\documentclass[conference]{IEEEtran}

\usepackage{cite}
\usepackage{amsmath,amssymb,amsfonts}
\usepackage{graphicx}
\usepackage{textcomp}
\usepackage{xcolor}
\usepackage{booktabs}
\usepackage{multirow}
\usepackage{comment}
\usepackage{url}
\usepackage{microtype}
\usepackage{balance}

\graphicspath{{figures/}}

\def\BibTeX{{\rm B\kern-.05em{\sc i\kern-.025em b}\kern-.08em
    T\kern-.1667em\lower.7ex\hbox{E}\kern-.125emX}}

\begin{document}

\title{Can LLMs Reason About Brand Ownership?\\
An Empirical Study of Domain Attribution Intelligence}

\author{
\IEEEauthorblockN{Fathima Mashood, Mohamed Nabeel}
\IEEEauthorblockA{fathima.mashood@snhu.edu, mmohamednabeel@nu.edu}
}

\maketitle

\begin{abstract}
When a new domain resembling a popular brand appears, defenders face a fundamental ambiguity: it may be an attacker-created squatting site for phishing, or it may be a domain the brand itself registered, either defensively, to block attackers, or legitimately, for a new product or service launch. Incorrectly flagging a brand-owned domain as malicious produces a false positive that harms end users and damages the brand's reputation. Resolving this ambiguity requires \emph{brand intelligence}: the ability to determine, at scale, whether a given domain belongs to a brand. Large language models (LLMs), with their broad knowledge of brand-domain relationships, offer a promising zero-configuration approach to this problem, but their reliability for brand intelligence tasks remains unknown. We present the first systematic empirical evaluation of LLM brand intelligence across three tasks: domain enumeration (Q1), open-ended brand attribution (Q2), and binary ownership classification (Q3). We evaluate four models, Gemini 2.5 Flash, Gemini 3.5 Flash, Claude Sonnet 4.5, and Claude Sonnet 4.6, across four retrieval settings (in-context, web search, WHOIS lookup, and combined) on 36 of the most-phished brands. Our results reveal a stark dichotomy: models achieve up to 82\% precision enumerating brand domains from memory alone, yet fail at ownership \emph{verification} without external tools, with macro-F1 at most 0.37 in ICL mode. WHOIS augmentation lifts Q3 macro-F1 by up to 0.65 points, yielding near-perfect precision ($\geq$0.99), dramatically reducing the false positive risk for defenders. We provide concrete recommendations for deploying LLMs in brand protection pipelines.
\end{abstract}


\section{Introduction}

Domain squatting, registering names visually or semantically similar to a trusted brand, is a scalable enabler of phishing, fraud, and credential theft. Attackers combine brand names with descriptive words (combosquatting~\cite{combosquatting2017}), embed them in TLS certificate SANs~\cite{embeddingsquatting:2019}, or craft long URL paths where the brand name appears only in a low-attention position (levelsquatting~\cite{Du2019TLDRHA}). The resulting pages convincingly impersonate brands such as PayPal, Google, and Microsoft to deceive end users~\cite{Tian:IMC2018:phishing}.

\subsection*{The False-Positive Problem in Brand Protection}

A critical and underappreciated challenge for defenders is distinguishing \emph{attacker-created} squatting domains from \emph{brand-created} domains that merely \emph{look} like squatting. Modern brands actively register large portfolios of brand-adjacent domains for legitimate purposes:

\begin{itemize}
  \item \textbf{Defensive registration:} Brands pre-emptively register domains that attackers are likely to squat on, typo variants, common TLDs, and combosquatting patterns, specifically to keep them out of attacker hands~\cite{brandshield2024,adjibi2025guardians}. A domain such as \texttt{google2.com} or \texttt{google-web.com} are owned by Google, not an attacker.
  \item \textbf{Product and service launches:} Brands register new domains for marketing campaigns, product microsites, regional services, and acquisitions. These (e.g. \texttt{microsoft365.com} or \texttt{paypal-prepaid.com}) may look squatting-like to an automated system that has no knowledge of the new registration.
\end{itemize}

Security systems that flag all brand-adjacent domains as malicious or phishing, without first checking brand ownership, generate significant false positives. A false positive on a brand-owned domain wrongly blocks users from reaching a legitimate service, erodes trust in the security tool, and can expose the defender to legal and reputational risk. \emph{Brand intelligence}, the ability to determine, at scale and in near-real-time, whether a given domain is owned by or affiliated with a known brand, is therefore a prerequisite step in any phishing or malicious-domain detection pipeline. Only after confirming that a domain is \emph{not} brand-owned should it be treated as a candidate for malicious classification.

\subsection*{Why Brand Intelligence Is Hard}

Maintaining accurate brand-domain knowledge is non-trivial. Brands continuously register new domains, acquire companies with their own domain portfolios, and silently retire old ones. No single public registry tracks all brand-owned domains. WHOIS registrant data can be privacy-proxied. Manual curation of brand domain lists does not scale across tens of thousands of brands at phishing-relevant freshness. What is needed is a system that can \emph{reason} about brand ownership for an arbitrary domain, including domains not previously seen in any brand database.

\subsection*{LLMs as Brand Intelligence Engines}

Large language models (LLMs) are trained on vast corpora that include brand announcements, WHOIS data discussions, corporate filings, and security research, giving them implicit knowledge of brand-domain relationships. In principle, an LLM can answer ``Is \texttt{google-payments.com} owned by Google?'' by drawing on this knowledge without a curated lookup table. Combined with real-time tools (web search, WHOIS lookup), LLMs could become scalable, zero-configuration brand intelligence engines capable of handling novel domains. Whether they are reliable enough for this role in practice is an open question.

Recent work has begun using LLMs for domain analysis~\cite{Chiba2025DomainLynxLL}, but no study has systematically evaluated the full brand intelligence pipeline, domain enumeration, brand attribution, and ownership verification, across multiple LLMs and retrieval settings. This paper fills that gap with a rigorous empirical evaluation designed to answer: \emph{can we rely on LLMs for brand intelligence, and under which conditions?}

\subsection*{Research Questions}

We study three complementary brand intelligence tasks:

\begin{itemize}
\item \textbf{Q1: Domain Enumeration:} Can an LLM enumerate the domains legitimately owned by a brand? (Building a brand domain inventory from scratch.)
\item \textbf{Q2: Brand Attribution:} Given an unseen domain (owned or squatting), can an LLM identify which brand it is associated with? (Contextualizing a domain for downstream analysis.)
\item \textbf{Q3: Ownership Verification:} Can an LLM determine whether a specific domain belongs to a specific brand? (The critical false-positive reduction step before malicious classification.)
\end{itemize}

\subsection*{Summary of Findings}

Our key findings are:
(1)~LLMs have strong memorised brand-domain knowledge, achieving mean Precision@50 up to 0.82 from memory alone, sufficient to seed a brand domain inventory.
(2)~Brand attribution (Q2) is generally accurate ($>$88\%), enabling coarse contextualisation of unknown domains.
(3)~Ownership verification (Q3), the step that directly reduces false positives,  \emph{requires} WHOIS tools: macro-F1 rises from at most 0.37 to over 0.85 when WHOIS lookup is available.
(4)~Web search augmentation is unreliable and sometimes degrades performance; WHOIS is the decisive tool.
(5)~Gemini models deliver near-comparable accuracy at 40--200$\times$ lower cost than Claude models, making them practical for production deployment.

\section{Related Work}

\subsection{Domain Squatting}

Domain squatting encompasses a family of techniques by which attackers register names visually or semantically similar to legitimate brands. Kintis et al.~\cite{combosquatting2017} characterise combosquatting at DNS scale, finding that nearly 60\% of abusive combosquatting domains persist for over 1,000 days. Roberts et al.~\cite{embeddingsquatting:2019} study target-embedding squatting in TLS certificates, where an attacker embeds a brand name as a subdomain (e.g., \texttt{google.evil.com}). Du et al.~\cite{Du2019TLDRHA} introduce levelsquatting, exploiting the tendency of users to read only the final segment of long URLs. Tian et al.~\cite{Tian:IMC2018:phishing} track elite phishing domains in the wild, noting that even short-lived domains can cause significant harm.

\subsection{Defensive Registration}

Brands respond to squatting by registering potential targets pre-emptively. Adjibi et al.~\cite{adjibi2025guardians} characterise defensive registration practices among Fortune 500 companies, revealing large variation in coverage. Benjamin et al.~\cite{brandshield2024} offer a complementary analysis focusing on how registration practices vary across brand-protection registrars and brand categories.

\subsection{LLMs for Domain Analysis}

Chiba et al.~\cite{Chiba2025DomainLynxLL} propose DomainLynx, which uses LLMs to detect squatting domains by reasoning about domain structure and brand names. That work focuses on \emph{detection} (binary squatting/benign) rather than \emph{attribution} (which brand does a domain belong to?), and evaluates only ICL-mode LLMs without external retrieval tools. In contrast, our work: (i)~isolates three distinct brand intelligence tasks with separate metrics, (ii)~systematically evaluates real-world tool augmentation (web search and WHOIS lookup) across four retrieval settings, (iii)~covers 36 brands spanning finance, technology, social media, and infrastructure, and (iv)~directly compares two LLM families (Gemini and Claude) across four model versions on the same eval dataset. To our knowledge, this is the first empirical study of LLM brand intelligence with tool-augmented retrieval settings and a purpose-built evaluation corpus.

\section{Dataset Construction}

\subsection{Brand Selection}

We focus on 36 brands selected from published lists of the most-phished targets, prioritising brands with high phishing volume and broad consumer exposure. The set spans finance (e.g., PayPal, Chase, Wells Fargo, Bank of America), technology (Google, Microsoft, Apple, Amazon), social media (Facebook, Instagram, Reddit, TikTok), cloud infrastructure (Cloudflare, Akamai, Fastly), and AI services (ChatGPT, Claude, OpenAI, Gemini). For brands with multiple primary domains (e.g., Alphabet's Google, YouTube, and Gemini), each primary domain is treated as a separate brand entry. For parent attribution, we use \texttt{alphabet.xyz} for Google-family brands and \texttt{meta.com} for Meta-family brands. Brands with privacy-proxied WHOIS registration (e.g., Cloudflare, Discord, Pinterest) are handled by falling back to registrar-only matching during dataset construction.

\subsection{Brand-Owned Domains}

For each brand we build a ground-truth set of owned domains via a six-step pipeline:

\begin{enumerate}
\item \textbf{Tranco WHOIS}: scan the Tranco top-1M list; accept domains whose registrant organisation matches the brand anchor via WHOIS.
\item \textbf{Crunchbase}: match company records by domain field with token-level brand keyword matching.
\item \textbf{DNStwist}: accept dnstwist-generated variants registered with brand-protection registrars (MarkMonitor, CSC, NOM-IQ, Com Laude, RegistrarSafe) or with matching registrant org.
\item \textbf{Search}: query DuckDuckGo for official brand sites; WHOIS-verify new candidates.
\item \textbf{TLD Sweep}: probe all 293 common TLDs with the brand name; accept via WHOIS match or reputed registrar.
\item \textbf{CT Logs}: query \texttt{crt.sh} for certificates where the SLD matches the brand name; WHOIS-verify new apex domains.
\end{enumerate}

Only registered apex domains (no subdomains) are retained. Registrant org matching uses word-level set comparison after stripping corporate suffixes. Brands with privacy-proxied WHOIS (e.g., Cloudflare, Discord) fall back to registrar-only matching.

\subsection{Squatting (Non-Brand) Domains}

Negative samples are drawn from dnstwist-generated variants for each brand that are \emph{not} in the owned set, supplemented with Tranco top-100 domains outside the brand portfolio. This yields up to 50 squatting/non-brand domains per brand.

\subsection{Evaluation Dataset}

The evaluation dataset contains, per brand: up to 50 owned domains and 50 non-brand/squatting domains. For Q3, these are directly used as binary classification instances. For Q2, we ask the model to name the associated brand for each domain (owned and squatting alike --- both should name the target brand). For Q1, the LLM is asked to enumerate domains, and its output is validated post-hoc against the owned set and WHOIS signals.

\section{Experimental Setup}

\subsection{Models}

We evaluate four state-of-the-art LLMs, all accessed via Google Cloud Vertex AI:

\begin{itemize}
\item \textbf{Gemini 2.5 Flash} (\texttt{gemini-2.5-flash})
\item \textbf{Gemini 3.5 Flash} (\texttt{gemini-3.5-flash}) 
\item \textbf{Claude Sonnet 4.5} (\texttt{claude-sonnet-4-5})
\item \textbf{Claude Sonnet 4.6} (\texttt{claude-sonnet-4-6})
\end{itemize}

\subsection{Retrieval Settings}

Each model is evaluated under four settings:

\begin{itemize}
\item \textbf{ICL}: in-context learning only; no external tools.
\item \textbf{Search}: LLM has access to a DuckDuckGo web search function tool (up to 2 tool-call rounds per query).
\item \textbf{WHOIS}: LLM has access to a WHOIS lookup function tool, backed by a third-party WHOIS API. Returns registrar, registrant organisation, and creation date.
\item \textbf{Search+WHOIS}: both tools available simultaneously.
\end{itemize}

All tool-augmented settings use a unified function-calling interface, limiting models to two tool-call rounds before forcing a text-only final response. Q1 uses only \emph{ICL} and \emph{Search} settings, since WHOIS requires knowing which domains to look up, domains are instead WHOIS-verified programmatically during validation.

\subsection{Prompts}

\begin{itemize}
\item \textbf{Q1:} ``List exactly 50 domains owned or officially operated by \{brand\}. Respond with a JSON array of exactly 50 apex domain names only, no explanation.''
\item \textbf{Q2:} ``What brand or company is the domain `\{domain\}' primarily associated with? Respond with only the brand name. If unknown, respond with `unknown'.''
\item \textbf{Q3:} ``Does the domain `\{domain\}' belong to or is it officially operated by \{brand\}? Answer with only `yes' or `no'.''
\end{itemize}

\subsection{Metrics}

\begin{itemize}
\item \textbf{Q1: Precision@50:} fraction of the 50 LLM-generated domains verified as legitimately brand-owned via multi-signal pipeline (owned-set membership, WHOIS org match, reputed registrar, DNS resolution check).
\item \textbf{Q2: Accuracy:} fraction of domains where the predicted brand matches the true associated brand (alias-normalised).
\item \textbf{Q3: Macro-F1:} mean F1 averaged over all 36 brands per model/setting. API errors are excluded and not counted as false negatives.
\end{itemize}

\section{Results}

\subsection{Q1: Domain Enumeration}

\begin{figure}[t]
  \centering
  \includegraphics[width=\columnwidth]{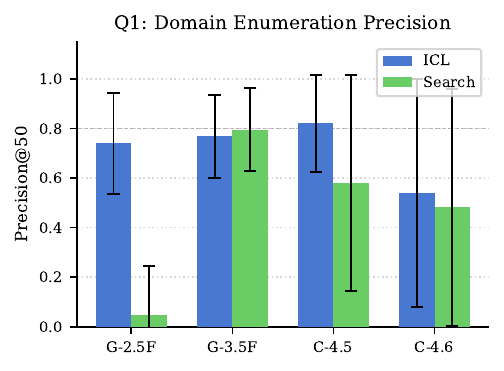}
  \caption{Q1 Precision@50 (mean $\pm$ std across 36 brands) for each model and retrieval setting. ICL = in-context learning; Search = DuckDuckGo web search tool.}
  \label{fig:q1-precision}
\end{figure}

\begin{figure}[t]
  \centering
  \includegraphics[width=\columnwidth]{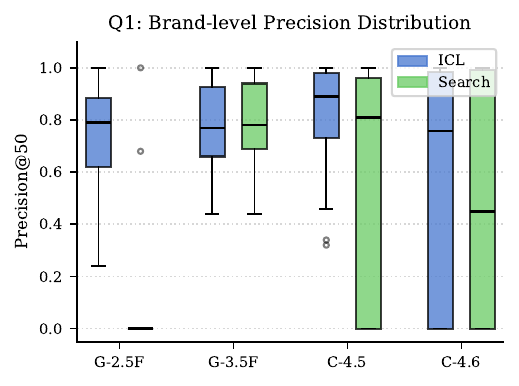}
  \caption{Distribution of brand-level Q1 Precision@50 per model--setting combination. Each box spans the interquartile range over 36 brands; whiskers extend to 1.5$\times$IQR.}
  \label{fig:q1-boxplot}
\end{figure}

Table~\ref{tab:q1} and Figures~\ref{fig:q1-precision}--\ref{fig:q1-boxplot} report Q1 results.

\begin{table}[t]
\caption{Q1 Precision@50 (mean $\pm$ std across 36 brands)}
\label{tab:q1}
\centering
\begin{tabular}{lcc}
\toprule
\textbf{Model} & \textbf{ICL} & \textbf{Search} \\
\midrule
Claude Sonnet 4.5  & \textbf{0.821} $\pm$ 0.196 & 0.580 $\pm$ 0.437 \\
Gemini 3.5 Flash   & 0.768 $\pm$ 0.169 & \textbf{0.795} $\pm$ 0.168 \\
Gemini 2.5 Flash   & 0.740 $\pm$ 0.203 & 0.047 $\pm$ 0.199 \\
Claude Sonnet 4.6  & 0.539 $\pm$ 0.459 & 0.481 $\pm$ 0.478 \\
\bottomrule
\end{tabular}
\end{table}

\textbf{ICL dominates for Claude; Search helps Gemini 3.5.}
Claude Sonnet 4.5 achieves the highest mean ICL precision (0.821), reflecting strong memorised knowledge of brand domain portfolios. Gemini 3.5 Flash is the only model to reliably benefit from web search (0.795 search vs.\ 0.768 ICL, std drops from 0.169 to 0.168), suggesting it effectively uses search results to enumerate domains beyond its training-data coverage.

\textbf{Gemini 2.5 Flash/Search collapses} to a mean precision of 0.047. When DuckDuckGo returns results (e.g., for Google), precision reaches 0.78; but for the majority of brands, particularly infrastructure and financial brands, the search tool timed out, and the model then generated plausible-sounding but non-existent domains.

\textbf{Claude Sonnet 4.6 shows extreme bimodal behaviour in ICL mode} (std=0.459): for well-known brands it reaches 1.00 precision, but for smaller-footprint brands (Fastly, Roblox, TikTok) it generates fabricated domains. The most common failure mode is inventing subdomains of the correct brand TLD (e.g., \texttt{fastly-cdn.net}) that do not exist.

\textbf{Hardest brands} across all ICL settings are infrastructure and security-focused brands with small public domain footprints: Fastly (0.32), Roblox (0.34), TikTok (0.46), and Cloudflare (0.60), where models hallucinate plausible but non-existent domains. Easiest brands (precision 1.00 for multiple models) are those with large, well-documented portfolios: Adobe, Amazon, Claude, and Gemini.

\subsection{Q2: Brand Attribution}

\begin{figure}[t]
  \centering
  \includegraphics[width=\columnwidth]{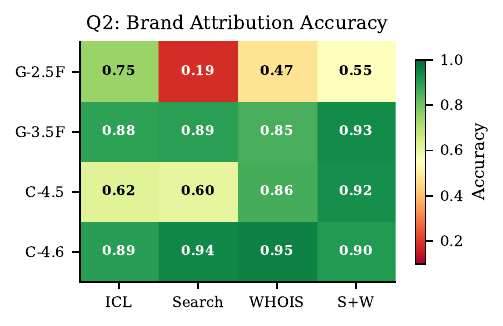}
  \caption{Q2 accuracy (fraction of domains correctly attributed to their associated brand) per model and retrieval setting. Warmer colours indicate higher accuracy.}
  \label{fig:q2-heatmap}
\end{figure}

\begin{table}[t]
\caption{Q2 Overall Accuracy (all domain categories; error records excluded)}
\label{tab:q2-overall}
\centering
\begin{tabular}{lcccc}
\toprule
\textbf{Model} & \textbf{ICL} & \textbf{Search} & \textbf{WHOIS} & \textbf{S+W} \\
\midrule
Claude Sonnet 4.6  & 0.892 & \textbf{0.940} & \textbf{0.950} & 0.900 \\
Gemini 3.5 Flash   & 0.885 & 0.891 & 0.854 & \textbf{0.933} \\
Gemini 2.5 Flash   & 0.746 & 0.187$^\dagger$ & 0.471 & 0.548 \\
Claude Sonnet 4.5  & 0.617 & 0.604 & 0.858 & 0.924 \\
\bottomrule
\multicolumn{5}{l}{$^\dagger$ Heavily impacted by DuckDuckGo failures ($\approx$80\% of calls).}
\end{tabular}
\end{table}

\begin{table}[t]
\caption{Q2 Accuracy by Domain Category. Model abbreviations: G-2.5F = Gemini 2.5 Flash; G-3.5F = Gemini 3.5 Flash; C-4.5 = Claude Sonnet 4.5; C-4.6 = Claude Sonnet 4.6. S+W = Search+WHOIS.}
\label{tab:q2-category}
\centering
\begin{tabular}{llcccc}
\toprule
\textbf{Category} & \textbf{Model} & \textbf{ICL} & \textbf{Srch} & \textbf{WHOIS} & \textbf{S+W} \\
\midrule
\multirow{4}{*}{Owned}
 & G-3.5F & \textbf{0.938} & 0.942 & \textbf{0.971} & \textbf{0.975} \\
 & C-4.6  & 0.895 & 0.938 & 0.979 & 0.961 \\
 & G-2.5F & 0.808 & 0.183 & 0.857 & 0.878 \\
 & C-4.5  & 0.663 & 0.581 & 0.961 & 0.974 \\
\midrule
\multirow{4}{*}{Squatting}
 & C-4.6  & 0.890 & \textbf{0.944} & 0.923 & 0.843 \\
 & G-3.5F & 0.836 & 0.846 & 0.749 & 0.889 \\
 & C-4.5  & 0.576 & 0.608 & 0.764 & 0.880 \\
 & G-2.5F & 0.695 & 0.164 & 0.124 & 0.251 \\
\bottomrule
\end{tabular}
\end{table}

Figure~\ref{fig:q2-heatmap} and Tables~\ref{tab:q2-overall}--\ref{tab:q2-category} summarise Q2 results.

\textbf{Claude Sonnet 4.6 leads brand attribution,} achieving the highest accuracy in 3 of 4 settings (up to 0.950 with WHOIS). It excels particularly on squatting domains in search mode (0.944), suggesting it can infer brand intent from domain string patterns even without lookup tools.

\textbf{WHOIS helps with owned domains, but confuses models on squatting.} WHOIS lookup confirms registrant identity for owned domains, boosting attribution accuracy. However, for squatting domains, the registrant reflects the squatter, not the target brand, causing model confusion. Gemini 3.5 Flash's squatting accuracy drops from 0.836 (ICL) to 0.749 (WHOIS), and Gemini 2.5 Flash plummets to 0.124, accepting the squatter's registrant as the brand.

\textbf{Claude Sonnet 4.5 shows a striking ICL gap:} its ICL accuracy (0.617) is 0.13 points below the next-weakest model (Gemini 2.5 Flash at 0.746), yet recovers to 0.924 with combined tools, a gain of 0.307 points, the largest Q2 improvement from retrieval augmentation. This suggests weak memorised brand-domain knowledge that is effectively compensated by tool access.

\textbf{Gemini 2.5 Flash/Search is catastrophically degraded} (0.187 overall) by DuckDuckGo failures: the tool returned no results for approximately 80\% of domains, causing the model to speculate with limited context.

\subsection{Q3: Binary Ownership Verification}

\begin{figure}[t]
  \centering
  \includegraphics[width=\columnwidth]{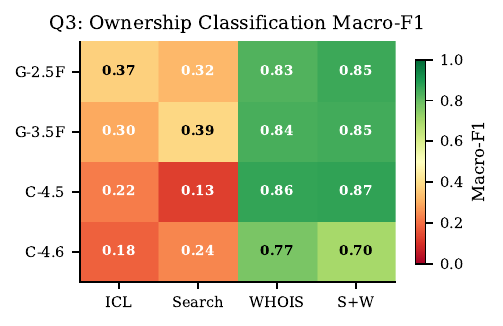}
  \caption{Q3 macro-F1 (averaged across 36 brands) per model and retrieval setting. WHOIS and Search+WHOIS settings show a dramatic improvement over ICL and Search.}
  \label{fig:q3-heatmap}
\end{figure}

\begin{figure}[t]
  \centering
  \includegraphics[width=\columnwidth]{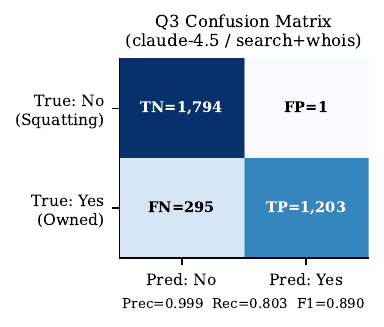}
  \caption{Q3 confusion matrix for Claude Sonnet 4.5 / Search+WHOIS --- the best-performing combination. TP=1,203; FP=1; TN=1,794; FN=295. Precision: 0.999; Recall: 0.803; F1: 0.891.}
  \label{fig:q3-confmat}
\end{figure}

\begin{table}[t]
\caption{Q3 Macro-F1 (mean F1 across 36 brands)}
\label{tab:q3-macro}
\centering
\begin{tabular}{lcccc}
\toprule
\textbf{Model} & \textbf{ICL} & \textbf{Search} & \textbf{WHOIS} & \textbf{S+W} \\
\midrule
Claude Sonnet 4.5  & 0.224 & 0.127 & 0.864 & \textbf{0.870} \\
Gemini 2.5 Flash   & 0.369 & 0.321 & 0.827 & 0.853 \\
Gemini 3.5 Flash   & 0.296 & 0.386 & 0.837 & 0.847 \\
Claude Sonnet 4.6  & 0.181 & 0.241 & 0.767 & 0.697 \\
\bottomrule
\end{tabular}
\end{table}

\begin{table}[t]
\caption{Q3 Aggregate Metrics (micro: all brands pooled). Abbreviations as in Table~\ref{tab:q2-category}. Spec. = Specificity (TNR). Selected rows shown.}
\label{tab:q3-micro}
\centering
\begin{tabular}{llccccc}
\toprule
\textbf{Model} & \textbf{Setting} & \textbf{Prec.} & \textbf{Rec.} & \textbf{F1} & \textbf{Spec.} \\
\midrule
C-4.5 & ICL        & 0.947 & 0.114 & 0.204 & 0.995 \\
C-4.5 & Search     & 0.974 & 0.078 & 0.145 & 0.999 \\
C-4.5 & WHOIS      & 0.999 & 0.792 & 0.884 & 1.000 \\
C-4.5 & S+W        & 0.999 & 0.803 & \textbf{0.891} & 0.999 \\
\midrule
C-4.6 & ICL        & 0.993 & 0.088 & 0.162 & 1.000 \\
C-4.6 & WHOIS      & 0.999 & 0.671 & 0.803 & 1.000 \\
C-4.6 & S+W        & 1.000 & 0.570 & 0.726 & 1.000 \\
\midrule
G-2.5F & ICL       & 0.837 & 0.258 & 0.395 & 0.957 \\
G-2.5F & WHOIS     & 0.994 & 0.758 & 0.860 & 0.996 \\
G-2.5F & S+W       & 0.987 & 0.803 & 0.886 & 0.991 \\
\midrule
G-3.5F & Search    & 0.987 & 0.256 & 0.406 & 0.997 \\
G-3.5F & WHOIS     & 0.996 & 0.751 & 0.856 & 0.997 \\
G-3.5F & S+W       & 0.997 & 0.763 & 0.865 & 0.998 \\
\bottomrule
\end{tabular}
\end{table}

Figures~\ref{fig:q3-heatmap}--\ref{fig:q3-confmat} and Tables~\ref{tab:q3-macro}--\ref{tab:q3-micro} report Q3 results.

\textbf{WHOIS is the decisive tool for Q3 and for false-positive reduction.}
Moving from ICL to WHOIS settings adds 0.47--0.65 macro-F1 points across all models. The task requires knowing who registered a domain, WHOIS answers this directly by revealing the registrant organisation. Without registration data, models cannot distinguish a brand's defensive registration from an attacker's impersonation based on domain strings alone. Practically, WHOIS augmentation reduces the false-negative rate from 74--92\% (ICL) to 20--33\% (WHOIS), meaning far fewer brand-owned domains are wrongly classified as ``not owned'', and consequently passed on to downstream detectors as potential threats.

\textbf{ICL and Search models are heavily biased toward ``no'' (not owned).}
In ICL mode, recall ranges from 0.08 to 0.26: models almost always predict ``no'' when uncertain. Precision is consequently near-perfect (0.84--0.99) since positive predictions are rare and confident. Critically, this bias creates a \emph{high downstream false-positive rate}: a recall of 0.11 means approximately 89\% of brand-owned domains are wrongly declared ``not owned'' by the model and would proceed to downstream malicious classification, exactly the false-positive risk that brand intelligence is meant to prevent. WHOIS lookup is the only tested setting that substantially corrects this bias.

\textbf{Search alone does not help Q3} and actually hurts recall for some models (Claude Sonnet 4.5: recall drops from 0.114 to 0.078). Failed DuckDuckGo queries leave models with even less context than ICL, reinforcing the conservative ``no'' default.

\textbf{Claude Sonnet 4.5 benefits most from WHOIS augmentation:} macro-F1 jumps from 0.224 to 0.870, the largest single tool-augmentation gain in our study. Claude Sonnet 4.6 with Search+WHOIS achieves perfect Q3 precision (1.000), it never wrongly clears a squatting domain as brand-owned, but its recall of 0.570 means 43\% of genuine brand-owned domains are still misclassified, leaving a substantial residual downstream false-positive risk. This precision--recall trade-off should inform deployment choices.

\textbf{Gemini models} achieve strong macro-F1 scores with WHOIS (0.827--0.853), and Gemini 2.5 Flash achieves the highest ICL recall (0.258), it is comparatively less conservative in the absence of tools, at a cost of precision (0.837 vs.\ 0.993 for Claude Sonnet 4.6).

\section{Discussion}

\subsection{Tool Benefit by Task}

Table~\ref{tab:tool-benefit} summarises the gain from the best tool setting over ICL.

\begin{table}[t]
\caption{Tool Benefit: Best Tool Setting vs.\ ICL Baseline. Q1 ICL outperforms search; negative sign indicates search degrades performance on average.}
\label{tab:tool-benefit}
\centering
\begin{tabular}{lccc}
\toprule
\textbf{Task} & \textbf{ICL Best} & \textbf{Best w/ Tools} & \textbf{Gain} \\
\midrule
Q1 (Precision@50) & 0.821 & 0.795 & $-$0.026 \\
Q2 (Accuracy)     & 0.892 & 0.950 & $+$0.058 \\
Q3 (Macro-F1)     & 0.369 & 0.870 & $+$0.501 \\
\bottomrule
\end{tabular}
\end{table}

WHOIS tools are transformative for Q3 and directly address the false-positive problem, while modestly improving Q2. For Q1, ICL from memory (Claude Sonnet 4.5, 0.821) actually \emph{outperforms} the best search-augmented result (Gemini 3.5 Flash, 0.795), and search catastrophically degrades performance for models that cannot handle search failures gracefully. This suggests that LLM training data encodes richer brand-domain knowledge than what real-time web search can surface reliably, particularly for brands with established domain portfolios.

\subsection{Recommended Brand Intelligence Pipeline}

For practitioners deploying LLMs as a brand intelligence pre-filter before malicious-domain classification, we recommend a three-stage pipeline:

\begin{enumerate}
  \item \textbf{Build an inventory (Q1):} Use Claude Sonnet 4.5 in ICL mode to enumerate known brand domains. Output feeds a WHOIS-validated allowlist. Do not use search tools for Q1 if the search tool is not reliable.
  \item \textbf{Attribute unseen domains (Q2):} Given a new domain, use Claude Sonnet 4.6 with WHOIS to identify the brand it is most likely associated with (95\% accuracy). Gemini 3.5 Flash / Search+WHOIS (0.933) is the cost-efficient alternative.
  \item \textbf{Verify ownership before flagging (Q3):} This is the critical false-positive gate. Use a WHOIS-augmented model to ask: ``Does this domain belong to Brand X?'' Only domains classified as ``not owned'' proceed to malicious classification. We recommend:
    \begin{itemize}
      \item \textbf{Balanced precision/recall:} Claude Sonnet 4.5 or Gemini 2.5 Flash with Search+WHOIS (macro-F1 $\approx$0.85--0.87, precision $\geq$0.987, recall $\geq$0.80).
      \item \textbf{Zero clearance of squatting domains:} Claude Sonnet 4.6 / WHOIS (precision=1.000, recall=0.671), no squatting domain is ever wrongly cleared, but 33\% of brand-owned domains still flow through to downstream analysis.
    \end{itemize}
\end{enumerate}

Without WHOIS augmentation (ICL or Search only), Q3 recall falls to 8--26\%, meaning 74--92\% of brand-owned domains would be sent downstream as potential threats, an unacceptable false-positive rate for any production system.

\subsection{Cost Considerations}

Claude models account for approximately 97\% of total experiment cost (\$58 of \$60) despite the same number of API calls, driven by 40--200$\times$ higher per-token pricing versus Gemini. For production deployments, Gemini models offer near-competitive performance at a fraction of the cost. The best performance-per-dollar combination across all 40 setups is \emph{Gemini 3.5 Flash / Search+WHOIS}, delivering 0.847 macro-F1 and 0.933 Q2 accuracy at approximately \$0.23 total experiment cost for 36 brands.

\subsection{Limitations}

\textbf{Search tool reliability.} DuckDuckGo timed out for approximately 5\% of all calls and up to 25\% of calls in search-heavy Q3 settings. Results for \emph{Search} and \emph{Search+WHOIS} settings should be interpreted with this caveat; a more reliable search API would likely improve these baselines.

\textbf{Eval set sampling.} Q2 and Q3 evaluations use a fixed random sample (seed 42) of 50 owned + 50 squatting domains per brand. Results may not generalise to all domain types, including recently registered squatting variants or novel TLD patterns. Further, they may not generalize to brands with a fewer domains.

\textbf{Tool round limit.} All models were restricted to two tool-call rounds per query. Deeper multi-step reasoning (e.g., search then WHOIS verify) might improve results in the Search+WHOIS setting.

\textbf{Knowledge cutoff and temporal staleness.} LLM training data has a fixed cutoff date. Recently registered brand domains, including new defensive registrations, product launches, or acquisitions, may not be in the model's training data, causing Q3 to wrongly classify them as ``not owned.'' This residual false-positive risk cannot be eliminated by ICL alone and underscores the importance of pairing LLM inference with live WHOIS lookups and periodic revalidation of brand domain inventories.

\section{Conclusion}

Defenders face a genuine dilemma when a new brand-adjacent domain appears: it may be an attacker's phishing site, or it may be a domain the brand registered for defensive purposes or a product launch. Flagging brand-owned domains as malicious produces false positives that harm users, erode trust in security tools, and carry legal risk. Brand intelligence, knowing whether a domain belongs to a brand, is the prerequisite step that resolves this ambiguity.

We conducted the first systematic empirical study of LLM brand intelligence across 36 brands, 4 models, and 4 retrieval settings. Our results draw a clear picture: LLMs carry strong memorised brand-domain knowledge (Q1 Precision@50 up to 0.82 from memory alone) and are generally reliable at brand attribution (Q2 accuracy up to 0.95). However, the task that matters most for false-positive reduction -- ownership verification (Q3) -- cannot be solved from memory: ICL-mode models misclassify 74--92\% of brand-owned domains as ``not owned,'' creating an unacceptable downstream false-positive rate. WHOIS augmentation is transformative: it lifts Q3 macro-F1 by up to 0.65 points and reduces the false-negative rate to 20--33\%, enabling near-precise clearance of brand-owned domains before any malicious classification is applied.

Web search augmentation, by contrast, is unreliable and sometimes degrades performance, it should not be used as the primary tool for ownership verification if the search tool is not relaible. Gemini models deliver near-competitive results at 40--200$\times$ lower cost than Claude, making them the practical choice for production deployments.

Our recommendation for practitioners is to deploy LLMs with live WHOIS lookups as a false-positive gate in phishing and malicious-domain detection pipelines, applied before, not after, malicious classification of brand-adjacent domains. Future work should explore fine-tuning on brand intelligence tasks, longer tool-call chains enabling richer multi-step reasoning, and integration with live certificate-transparency feeds to handle newly registered domains that fall outside LLM training-data cutoffs.

\balance
\bibliographystyle{IEEEtran}
\bibliography{ref}

\end{document}